\documentclass[aps,twocolumn]{revtex4-1}
\usepackage{hyperref}
\hypersetup{colorlinks=true, citecolor=blue, urlcolor=blue, linkcolor=blue}

\pdfoutput=1
\usepackage[utf8]{inputenc}
\usepackage[T1]{fontenc}
\usepackage{calc}
\usepackage{graphics,graphicx,amsfonts,amsmath,amsbsy,amssymb,color}
\usepackage{bm}
\usepackage{subfigure}
\usepackage{threeparttable}

\begin{document}

\title{Towards efficient and accurate \emph{ab initio} solutions to periodic systems via
transcorrelation and coupled cluster theory}

\author{Ke Liao}
\email{ke.liao.whu@gmail.com}
\affiliation{
Max Planck Institute for Solid State Research, Heisenbergstrasse 1, 70569 Stuttgart, Germany
}

\author{Thomas Schraivogel}
\affiliation{
Max Planck Institute for Solid State Research, Heisenbergstrasse 1, 70569 Stuttgart, Germany
}

\author{Hongjun Luo}
\affiliation{
Max Planck Institute for Solid State Research, Heisenbergstrasse 1, 70569 Stuttgart, Germany
}

\author{Daniel Kats}
\email{d.kats@fkf.mpg.de}
\affiliation{
Max Planck Institute for Solid State Research, Heisenbergstrasse 1, 70569 Stuttgart, Germany
}

\author{Ali Alavi}
\email{a.alavi@fkf.mpg.de}
\affiliation{
Max Planck Institute for Solid State Research, Heisenbergstrasse 1, 70569 Stuttgart, Germany
}

\affiliation{
Yusuf Hamied Department of Chemistry, University of Cambridge, Lensfield Road, Cambridge CB2 1EW, United Kingdom
}

\begin{abstract}
    We propose a streamlined combination scheme of the transcorrelation (TC) and 
    coupled cluster (CC) theory, which
    not only increases the convergence rate with respect to the basis set, but also
    extends the applicability of the lowest order CC approximations to strongly 
    correlated regimes in 
    the three dimensional uniform electron gas (3D UEG). 
    With the correct physical insights built into the correlator used in TC, 
    highly accurate ground state energies with errors $\leq 0.001 $ a.u./electron 
    relative to the state-of-the-art
    quantum Monte Carlo results can be obtained across a wide range of densities.
    The greatly improved efficiency and accuracy of our methods hold great promise for strongly
    correlated solids where many other methods fail.
\end{abstract}

\maketitle

\section{Introduction}
The coupled cluster (CC) methodologies~\cite{cizek1966,Cizek1971a,Cizek1980} at the 
level of singles and doubles (CCSD) and perturbative triples (CCSD(T))~\cite{Raghavachari:89} 
have become the \emph{de facto} standard of single-reference 
\emph{ab initio} quantum chemistry, and can be applied to systems consisting of hundreds 
of electrons~\cite{riplingerSparse2016,schmitzPerturbative2016,schwilkScalable2017,nagyOptimization2018a}.
In the past few years, these methods have also shown promise in 
applications to the solid state~\cite{Booth2013,Liao2016a,McClain2017,Gruber2018,Liao2019,Gao2020a}, although significant challenges remain before 
they can be routinely applied, as for example density functional theories are. 
On the one hand, because of quite steep computational scaling (${\cal O}(N^{6})$ and ${\cal O}(N^{7})$ for 
CCSD and CCSD(T) respectively), it is desirable to keep the methods at the lowest 
possible CC level, namely CCSD, whilst maintaining accuracy. The more accurate
CCSD(T), as a perturbative correction to CCSD, additionally fails for metals~\cite{Shepherd2013a}. 
It is also desirable that the CC methods can be extended to more strongly 
correlated systems, where the single reference nature of these approximations 
breaks down. There have been various attempts to develop modified CCSD methods with a higher accuracy 
for weakly \cite{Meyer:71,nooijen_orbital_2006, Neese:09,huntington_pccsd:_2010} and strongly \cite{paldus_approximate_1984,piecuch_solution_1991,bartlett_addition_2006,robinson_approximate_2011,small_coupled_2014} 
correlated systems. 
The distinguishable cluster singles and doubles (DCSD) \cite{Kats2013c,Kats2014} is one such method,
%[i think this part is too technical for an introduction. I have shifted it to a later part Ali]
%which can be derived as a small modification of the CCSD amplitude equations by neglecting inter-cluster exchange diagrams and
%ensuring the particle-hole symmetry and exactness for two electrons, and 
which has shown promise in improving CCSD in weakly and  strongly correlated molecular systems~\cite{kats_accurate_2015,rishi_excited_2017,tsatsoulis_comparison_2017}.
%and here we also investigate its performance in the uniform electron gas context of transcorrelation.

In a separate development, there has been renewed interest in so-called 
transcorrelated (TC) methods~\cite{Boys1969,Boys1969b,Boys1969c,Hino2001,Hino2002,Umezawa2004,Sakuma2006,Ochi2015b,Ochi2016,Ochi2017,Luo2018,Jeszenszki2018,Dobrautz2019,Cohen2019,Jeszenszki2020}, 
based on Jastrow factorisation of the electronic 
wavefunction, which result in effective similarity transformed (ST) Hamiltonians~\cite{Luo2018,Cohen2019}. 
Although TC methods were originally proposed as a way to accelerate basis set 
convergence in electronic wavefunctions, it has become apparent that such  
similarity transformations can also be extremely helpful in the context of 
strongly correlated systems. For example, in the repulsive 2D Fermi-Hubbard model, 
it was found that, with a suitable Gutzwiller correlator, 
extremely compact forms 
of ground state right eigenvectors of the ST Hubbard Hamiltonian could be obtained~\cite{Dobrautz2019}, 
dominated by the Hartree-Fock (HF) determinant. Since single-reference coupled cluster 
methods work best when the wavefunction is dominated by the HF 
determinant, and furthermore, since the coupled cluster method can itself be cast 
in terms of a similarity transformation of the Hamiltonian, it is natural to ask if the two 
concepts - coupled cluster and transcorrelation - can be usefully combined into 
a single framework, whereby the compactification generated by the 
TC method is exploited by the CC method, to extend its range of applicability 
into more strongly correlated systems. The purpose of this paper 
is to report such an investigation, applied to the uniform electron gas (UEG), 
over a broad range of densities.  

The 3D UEG model assumes that the background is evenly and positively charged, and that 
the electrons interact with each other via the Coulomb interaction. As simple as it is, the UEG possesses 
an intricate phase diagram~\cite{Wigner1934a,Ceperley1980}, which can only be accurately described by theories that perform consistently well over a broad range of densities.
Historically, the UEG model has also played an important role in the development of many useful approximations. 
For example, several successful local and gradient-corrected density functionals~\cite{Kohn1965b,Perdew1981,Perdew1996} are based on the UEG; the random phase approximation (RPA)~\cite{Bohm1953,Gell-Mann1957b} was 
developed in a pursuit of understanding metals using the UEG as a model. 
In recent years, the UEG has attracted studies from various highly accurate \emph{ab initio} methods and
spurred the development of several 
new methods~\cite{Umezawa2004,Shepherd2012,Shepherd2012h,Shepherd2013a,Shepherd2014a,Neufeld2017,Ruggeri2018,Luo2018,Irmler2019a}.
 
When applying CC to the UEG, we work in a plane wave basis; 
momentum conservation then excludes all 
single excitations from the CC ansatz, 
greatly simplifying the resulting amplitude equations. 
As a result, the TC Hamiltonian can be treated with relative ease, 
allowing us to investigate whether
the CC method can be beneficially applied to the TC Hamiltonian. 
We will investigate the CCD and DCD approximations, 
%$\hat{T}\approx \hat{T}_2 = \sum_{ijab} t^{ab}_{ij} \hat{t}_{ij}^{ab}$ 
in the 
context of the TC Hamiltonian, and show that with a suitable form
of the correlator, the basis set convergence can be greatly accelerated 
(as expected), but in addition highly accurate energies can be obtained 
across a broad range of densities $0.5\le r_s \le 50$, covering both the weakly 
and strongly correlated regimes. This gives us confidence 
that the method, once suitably generalised to real systems (which will need to 
include the singles contribution), will allow a highly accurate
yet efficient methodology for the solid state.

In the rest part of this paper, we review the UEG model, CC and TC theories in Sec.~\ref{sub:UEG},
~\ref{sub:CC} and~\ref{sub:TC}, respectively; in Sec.~\ref{sub:Approx} we discuss the 
important approximations made to the TC Hamiltonian; we demonstrate our scheme for choosing
the optimal parameters in the correlator in Sec.~\ref{sec:Corr}; we showcase and discuss
our TC-CCD/DCD results in comparison with benchmark data in Sec.~\ref{sec:Results}; and
finally we conclude the paper in Sec.~\ref{sec:Con} with some outlooks for future directions.

\section{Theory}
\label{sec:Theory}
\subsection{Three Dimensional Uniform Electron Gas}
\label{sub:UEG}
The 3D UEG is the simplest model for realistic periodic solids, 
of which the
Hamiltonian in real space reads
\begin{equation}
  \hat{H} = -\sum_{i}\frac{1}{2}\nabla^2_i 
  + \sum_{i\neq j}\frac{1}{|{\bf r_i-r_j}|} + {\rm const.},
\end{equation}
where the const. includes the interactions between electrons and the 
homogenous positive background charge, and the interactions 
between the electrons and their own periodic images, which is
termed as the Madelung constant and will disappear as the size of the
simulation cell goes to infinity. Atomic units are used to simplify the
equations.
When plane wave basis functions and a simple cubic simulation cell 
of volume $\Omega=L^3$ are used, we can 
reformulate the Hamiltonian in a second-quantized form,
\begin{equation}
  \hat{H} = \sum_{p}\sum_{\sigma}\frac{1}{2}{\bf k}_p^2a^{\dagger}_{p,\sigma}a_{p,\sigma}+ 
  \frac{1}{2}\sum_{pqrs}\sum_{\sigma\sigma'} V^{rs}_{pq}
  a^{\dagger}_{{p},\sigma}a^{\dagger}_{{q},\sigma'}a_{{s},\sigma'}a_{{r},\sigma},
\end{equation}
where for simplicity we use $p,q,r,s\dots$ indices as a compact form for 
the general momentum (plane wave basis function) 
indices ${\bf k}_p,{\bf k}_q,{\bf k}_r,{\bf k}_s\dots$ 
and hereon we use the two terms plane wave basis function and orbital equivalently.
We stress that due to momentum conservation, i.e.~${\bf k}\equiv{\bf k}_r-{\bf k}_p={\bf k}_q-{\bf k}_s$, there are only three free indices
among $pqrs$, and 
the interactions with the homogenous positive background charge 
are cancelled by the divergent Coulomb potential at $k=0$, which is defined as $V^{rs}_{pq}\equiv V({\bf k})=\frac{4\pi}{\Omega{\bf k}^2}$.
We also ignore the Madelung contribution in the Hamiltonian which 
can be added posteriorly to the ground state energy. 
%The operators $\hat{E}^{p}_{q}=\sum_{\sigma}\hat a^\dagger_{p\sigma}\hat a_{q\sigma}$ and 
%$\hat E_{rs}^{pq}=\sum_{\sigma}\hat a^\dagger_{p\sigma} \hat{E}^{q}_{s} \hat a_{r\sigma}$ 
%are the usual singlet second-quantization excitation operators. 
The electron density of the system can be described by the
Wigner-Seitz radius $r_s=\left(\frac{3}{4\pi N}\right)^{1/3}L$, where $N$ is the number of electrons.

\subsection{Coupled/Distinguishable Cluster Doubles}
\label{sub:CC}
In the CC ansatz, we let the many-electron ground state wavefunction to be 
\begin{equation}
\Phi= e^{\hat{T}} \Phi_{0}
\end{equation}
where $\Phi_0$ is the HF wavefunction and $\hat{T}$ is a cluster 
operator. In the case of the UEG, we work in a plane wave basis and  $\Phi_0$ 
is given by the Fermi sphere. 
We will investigate the CCD approximation and its distinguishable variant 
(DCD)~\cite{Kats2013a,Kats2013c}, which is based on a modification of the CCD amplitude equations by neglecting inter-cluster exchange diagrams and
ensuring the particle-hole symmetry and exactness for two electrons.
Alternatively, DCD can be derived from screened Coulomb considerations~\cite{katsDistinguishable2016}.
We start with the canonical CCD and later highlight the
differences between CCD and DCD.

In CCD, the full cluster operator is approximated by 
the doubles excitations only,
\begin{equation}
  \hat{T}\approx \hat{T}_2 = \frac{1}{2}\sum_{ijab} T_{ab}^{ij}\sum_{\sigma\sigma'}
  a^{\dagger}_{{a},\sigma}a^{\dagger}_{{b},\sigma'}a_{{j},\sigma'}a_{{i},\sigma},
\end{equation}
where $T^{ij}_{ab}$ are the doubles amplitudes. Following convention, 
we use $i,j,k\dots$ and $a,b,c\dots$ to represent occupied and unoccupied 
orbitals in the reference determinant, respectively.
Again, the momentum conservation ensures that only 3 indices
of the amplitude tensor are free, saving a great deal in storing them in 
the computer memory. 

The $T_2$ amplitudes are obtained by solving the projective doubles amplitude equations,
\begin{equation}
  \left<\Phi_{ij}^{ab}\left|e^{-\hat{T}_2}\hat{H}e^{\hat{T}_2}\right|\Phi_0\right> = 0,
  \label{eq:amps}
\end{equation}
where $\Phi_{ij}^{ab}$ are doubly substituted determinants.
%where $E_{\rm HF}=\left<\Phi_0\right|\hat{H}\left|\Phi_0\right>$. \
To be specific, a functional form of the residual, which unifies CCD and DCD for closed shell systems,
can be written as
\begin{equation}
  \begin{aligned}
    R^{ij}_{ab} = &V_{ab}^{ij}+V_{ab}^{cd}T_{cd}^{ij}+I_{kl}^{ij}T_{ab}^{kl} + X_{al}^{cj}T_{cb}^{il}+ \tilde{T}_{ac}^{ik} V_{kl}^{cd}\tilde{T}_{db}^{lj} \\
          & + \mathcal{\hat{P}}(ia;jb)\left[x_a^cT_{cb}^{ij}-x_k^iT_{ab}^{kj} + \chi_{al}^{ci}(T_{bc}^{lj}-T_{cb}^{lj})\right.\\
          &\left.-V_{ka}^{ic}T_{cb}^{kj}-V_{kb}^{ic}T_{ac}^{kj}+\tilde{T}_{ac}^{ik}V_{kb}^{cj}\right],
  \end{aligned}
\end{equation}
where we define the permutation operator $\mathcal{\hat{P}}(ia;jb)T_{ab}^{ij}\equiv T_{ab}^{ij}+T_{ba}^{ji}$
and the following intermediates,
\begin{align}
  I_{kl}^{ij} &= \begin{cases}
                V_{kl}^{ij} + V_{kl}^{cd}T_{cd}^{ij},& \quad\text{CCD}\\
                V_{kl}^{ij},& \quad\text{DCD}
  \end{cases}\\
  X_{al}^{cj} &= \begin{cases}
                V_{kl}^{cd}T_{ad}^{kj},& \quad\text{CCD}\\
                0,& \quad\text{DCD}
  \end{cases}\\
  x_a^c &= \begin{cases}
        f_a^c - \tilde{T}_{ad}^{kl}V_{lk}^{dc},& \quad\text{CCD}\\
        f_a^c - \frac{1}{2}\tilde{T}_{ad}^{kl}V_{lk}^{dc},& \quad\text{DCD}
  \end{cases}\\
  x_k^i &= \begin{cases}
        f_k^i + \tilde{T}_{cd}^{il}V_{lk}^{dc},& \quad\text{CCD}\\
        f_k^i + \frac{1}{2}\tilde{T}_{cd}^{il}V_{lk}^{dc},& \quad\text{DCD}
  \end{cases}\\
  \chi_{al}^{ci} &= \begin{cases}
    V_{kl}^{cd}T_{da}^{ki},& \quad\text{CCD}\\
    0,& \quad\text{DCD}
    \end{cases}\\
    \tilde{T}_{ab}^{ij} &= 2T_{ab}^{ij}-T_{ba}^{ij}.
\end{align}
We note that in this case the Fock matrix $f^p_q$ is diagonal, with the
diagonal elements being the orbital energies $\epsilon_p$. 
A straightforward way to update the $T_2$ amplitudes at iteration $n+1$  will be
\begin{align}
  \Delta_{ab}^{ij} &= \frac{R_{ab}^{ij}}{\epsilon_i+\epsilon_j-\epsilon_a-\epsilon_b},\\
  T_{ab}^{ij} (n+1)&=  T_{ab}^{ij}(n) + \Delta_{ab}^{ij}.
\end{align}
Of course, more advanced iterative schemes can be used,~e.g.~DIIS~\cite{Pulay1980,Pulay1982},
to accelerate convergence rate.

Using the converged $T_2$ amplitudes, the correlation energy is expressed as
\begin{equation}
  E_{\rm c} = \tilde T_{ab}^{ij} V_{ij}^{ab},
\end{equation}
and the total energy is 
\begin{equation}
  E = E_{\rm HF} + E_{\rm c},
\end{equation}
where $E_{\rm HF}=\left<\Phi_0\right|\hat{H}\left|\Phi_0\right>$.

\subsection{Transcorrelation}
\label{sub:TC}
In the transcorrelation framework the 
many-electron wavefunction is written as
\begin{equation}
  \Psi = e^{\hat{\tau}} \Phi
\end{equation}
where $\hat{\tau}=\frac{1}{2}\sum_{i\neq j}u({\bf r_i,r_j})$ is a correlator 
consisting of pair correlations $u({\bf r_i,r_j})$, whose form will 
be discussed later. $\Phi$ should satisfy the similarity-transformed 
eigenvalue equation
\begin{equation}
  \hat{H}_{\rm tc} \Phi = E \Phi, \quad \hat{H}_{\rm tc} = e^{-\hat{\tau}} \hat{H} e^{\hat{\tau}}.
\end{equation}
It is worth pointing out that at this stage, no approximations have been made,
and the spectra $E$ of $\hat{H}_{\rm tc}$ are the same as of the original Hamiltonian.
As shown in Ref.~{\cite{Luo2018}}, the second quantized form of the $\hat{H}_{\rm tc}$ in a 
plane wave basis is
\begin{equation}
\begin{aligned}
  \hat{H}_{\rm tc} &= \hat{H}+
                  \frac{1}{2}\sum_{\sigma\sigma'}\sum_{pqrs}
                  {\omega}_{pq}^{rs}a^{\dagger}_{{p},\sigma}a^{\dagger}_{{q},\sigma'}
                  a_{{s},\sigma'}a_{{r},\sigma} \\
                  &+ \frac{1}{2}\sum_{\sigma\sigma'\sigma''}\sum_{pqorst}
                  {\omega}^{rst}_{pqo}
                  a^{\dagger}_{{p},\sigma}a^{\dagger}_{{q},\sigma'}a^{\dagger}_{{o},\sigma''}  
                  a_{{ t},\sigma''}a_{{ s},\sigma'}a_{{ r},\sigma},
\end{aligned}
\end{equation}
where momentum conservation requires
  ${\bf k} \equiv {\bf k}_r-{\bf k}_p$,
  ${\bf k'} \equiv {\bf k}_q-{\bf k}_s$ and
  ${\bf k}_o = {\bf k}_t+{\bf k-k'}$, and we define
\begin{align}
  {\omega}^{rs}_{pq} &= \frac{1}{\Omega}\left[k^2\tilde{u}({\bf k}) - ({\bf k}_r-{\bf k}_s)\cdot {\bf k} \tilde{u}({\bf k})\right]\\
  \nonumber
  &\quad+\frac{1}{\Omega}\sum_{\bf k'}({\bf k-k'})\cdot {\bf k'} \tilde{u}({\bf k-k'})\tilde{u}({\bf k'}), \\
  {\omega}^{rst}_{pqo} &= -\frac{1}{\Omega^2}\tilde{u}({\bf k})\tilde{u}({\bf k'}){\bf k}\cdot{\bf k'}.
\end{align}
The TC Hamiltonian
has additional 2-body and 3-body interactions.
Due to one of the additional 2-body interactions, the TC Hamiltonian is non-hermitian.
This fact can pose some difficulties for variational methods, but not so for projection methods such as
full configuration interaction Monte Carlo (FCIQMC)~\cite{Booth2009a,Booth2013} and CC. 

\subsection{Approximations to the 3-body operator}
\label{sub:Approx}
The additional 3-body operator when treated without
approximations will increase the computational scaling of CCD or DCD from $N^6$ to
$N^7$. To seek a good balance between the 
computational cost and the accuracy, 
we include only up to effective 2-body operators arising from 
normal-ordering the 3-body operator.
In this approximation, only the normal-ordered 
3-body operator is excluded. 
We can justify this approximation by analogy to the HF approximation, which
constructs a mean-field solution by including only the single and double contractions 
from the Coulomb operator. In cases where the mean-field approximation is reasonably good, the contribution of the missing normal-ordered Coulomb operator is small,
compared to the single and double contractions. 
In contrast to the HF approximation, the
parameters in the correlator in general 
allow a tuning of the strength of the 
missing normal-ordered 3-body operator, which we will discuss in the next section.

In general, we can write our approximated
Hamiltonian as 
\begin{equation}
    \begin{aligned}
        \hat{H}_{\rm tc} &=  \tilde{E}_{\rm HF} + \sum_{\sigma}\sum_{ p} \tilde{\epsilon}_{ p} \{a^{\dagger}_{{ p},\sigma}a_{{ p},\sigma}\}\\
        &+\frac{1}{2} \sum_{\sigma\sigma'}\sum_{ pqrs}\tilde{V}^{rs}_{pq}
        \{a^{\dagger}_{{ p},\sigma}a^{\dagger}_{{ q},\sigma'}a_{{ s},\sigma'}a_{{ r},\sigma}\}\\
        &+ E_{\rm T} + \sum_{\sigma}\sum_{ p} \tilde{\omega}_{ p} \{a^{\dagger}_{{ p},\sigma}a_{{ p},\sigma}\} \\
        &+ \frac{1}{2}\sum_{\sigma\sigma'}\sum_{pqrs}\tilde{w}^{rs}_{pq}\{a^{\dagger}_{{ p},\sigma}a^{\dagger}_{{ q},\sigma'}a_{{ s},\sigma'}a_{{ r},\sigma}\},
    \end{aligned}
    \label{equ:normal_ordered_tc_ham}
\end{equation}
where $E_T$ refers
to the triply-contracted 3-body operator contribution, $\tilde{\omega}_{p}$ is the 
doubly-contracted 3-body integral and 
$\tilde{\omega}^{rs}_{pq}$ is the singly-contracted 3-body integral. The curly
brackets indicate that the operators are normal-ordered with respect to the 
HF vacuum (Fermi sphere). We emphasize that in Eq.~(\ref{equ:normal_ordered_tc_ham})
$\tilde{E}_{\rm HF}$ and $\tilde{\epsilon}_p$ are calculated now
with the modified 2-body integrals $\tilde{V}_{pq}^{rs}=w^{rs}_{pq}+V^{rs}_{pq}$.
For clarity, we outline the procedures for our TC-CCD/DCD framework. 
\begin{enumerate}
    \item Evaluating $\omega^{rs}_{pq}$ and $V^{rs}_{pq}$ and combining them into $\tilde{V}_{pq}^{rs}\leftarrow w^{rs}_{pq}+V^{rs}_{pq}$;
    \item Calculating $\tilde{\epsilon}_p = \frac{{\bf k}_p^2}{2} + \sum_i(2\tilde{V}^{pi}_{pi}-\tilde{V}^{pi}_{ip})$; 
    \item Calculating $\tilde{E}_{\rm HF} = 2\sum_i^{N/2}\tilde{\epsilon}_i-\sum_{ij}(2\tilde{V}^{ij}_{ij}-\tilde{V}^{ij}_{ji})$ and $E_{\rm T}$;
    \item Evaluating $\tilde{\omega}_{ p}$, and defining $\epsilon_p \leftarrow \tilde{\epsilon}_p+\tilde{\omega}_{ p}$;
    \item Evaluating the singly-contracted 3-body integral $\tilde{w}^{rs}_{pq}$ and redefining
    $V^{rs}_{pq}\leftarrow \tilde{V}_{pq}^{rs}+\tilde{w}^{rs}_{pq}$;
    \item Solving the usual CCD/DCD amplitude equations using $\epsilon_p$ and $V^{rs}_{pq}$ for $T_2$ and obtaining $E_{\rm c}$;
    \item The total energy is $E=\tilde{E}_{\rm HF}+E_{\rm T}+E_{\rm c}$.
\end{enumerate}

For details on the mathematical expressions for the contractions
of the 3-body operator, we refer to the Supplementary Information.

\section{Choice and optimization of the correlator}
\label{sec:Corr}
Past experience with the TC method has shown that the form of the correlator 
$\hat{\tau}$ is of extreme importance in the TC method, otherwise the benefit
of transcorrelation is lost -- $\Phi$ can be simpler than $\Psi$ only if the 
correlator captures the correct physics of the pair correlations in the system.    
An inappropriate correlator can actually lead to a harder problem than the 
original Schr\"odinger equation. In our previous study of the exact TC method in 
the UEG, we proposed a form of the correlator (shown below) which was found to 
work successfully in accelerating convergence to the basis set limit, without 
changing the correlation that could be captured with the basis set by a FCI 
level $\Phi$ function.  In the present study, since we will be approximating 
the $\Phi$ with the CC ansatz, we additionally require the 
correlator $\hat{\tau}$ to capture some of the correlation inside the Hilbert space. 

\begin{figure}[th!]
  \centering
  \includegraphics[width=\linewidth]{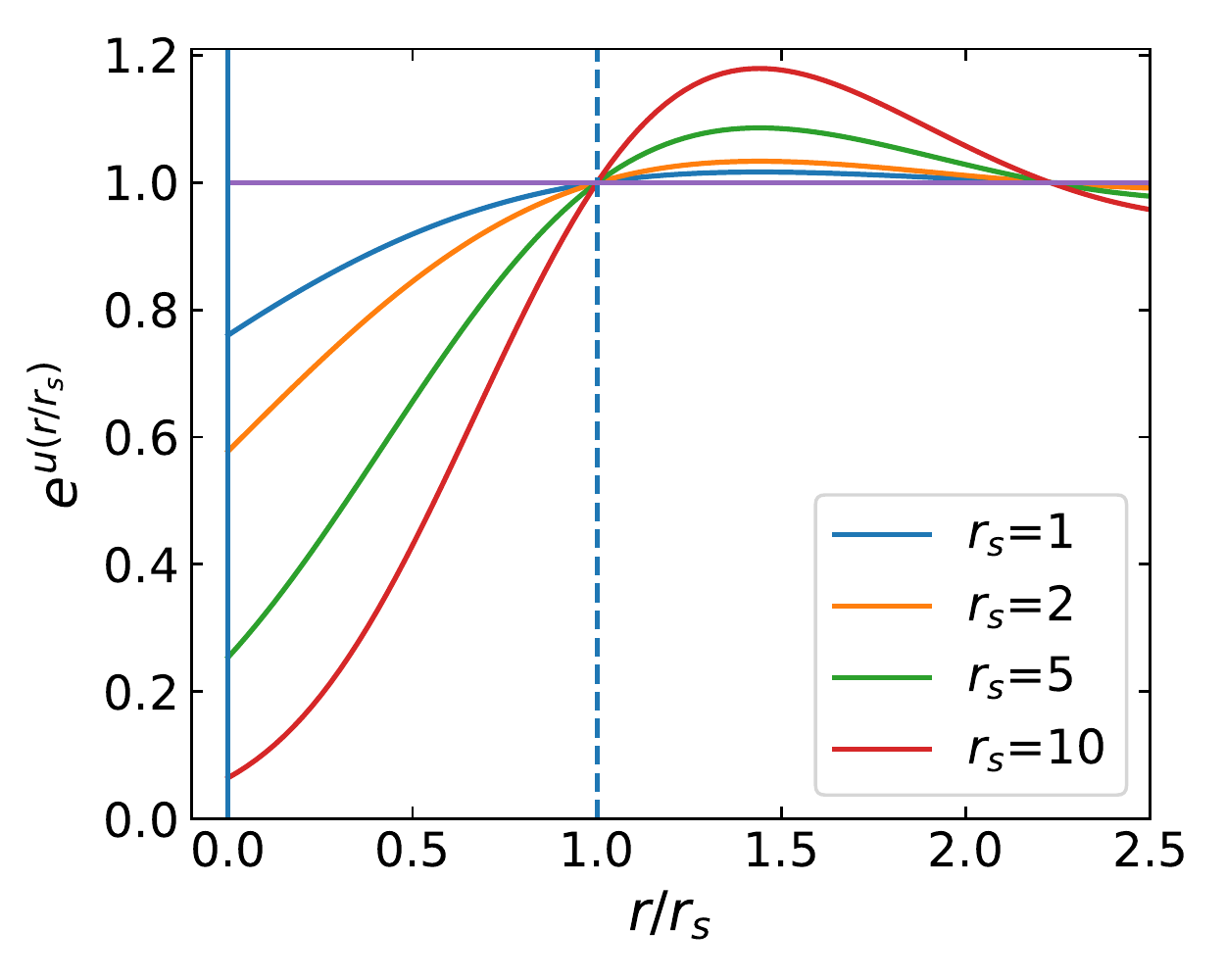}
  \caption{A sketch of the Jastrow factor with the proposed correlator as the 
  exponent.
  }
  \label{fig:correlator}
\end{figure}

Here we propose a physically motivated correlator that mimics the
behavior of the correlation hole between two unlike-spin electrons as $r_s$ varies in 3D UEG.
The correlation hole can be examined by the pair-correlation function $g(r_{12})$
in real space, as studied in Ref.~\cite{Gori-Giorgi2000}, which shows that the correlation hole between two unlike-spin electrons
grows deeper and wider as the Wigner-Seitz radius $r_s$ increases or as the electron 
density decreases. Fig.~\ref{fig:correlator} provides a sketch of the Jastrow factor with 
our proposed correlator $u(r_{12})$ as the exponent, which captures the
desired behavior. 
We point out that
the functional form of this correlator, which reads in real and reciprocal space
respectively as
%\begin{equation}
  \begin{align}
    u(r) &= -\frac{r}{\pi}\left({\rm si}(k_c r)+\frac{{\rm cos}(k_cr)}{k_cr}+\frac{{\rm sin}(k_cr)}{(k_cr)^2}\right), \label{equ:ReCorrelator}\\
    \tilde{u}({\bf k}) &= \begin{cases} -\frac{4\pi}{k^4}, \quad |{\bf k}| > k_c,\\
                                    0, \quad |{\bf k}| \leq k_c,
                        \end{cases}
  \end{align}
%\end{equation}
where ${\rm si}(x) = -\int_x^{\infty} \frac{\sin(t)}{t}\mathrm dt$, was first reported in Ref.~\cite{Luo2018} to satisfy 
the cusp condition between two electrons with opposite spins at short inter-electron distance and its influence is reduced to
nonexistence as the complete basis set (CBS) limit is reached. This was done by choosing 
$k_c$ to be the same as the plane wave cutoff momentum, $k_{\rm F}$, which defines
how many plane waves are included as basis functions.
In contrast, to mimic the behavior of the pair-correlation function,
as a first attempt in the present study  
we choose the parameter in this correlator such that the first nonzero root
of Eq.~(\ref{equ:ReCorrelator})
is fixed to be at
$r_s$, irrespective of the basis set. This is achieved by setting
\begin{equation}
  k_c = \frac{R_1}{r_s},
\end{equation}
where $R_1\approx2.322502989$.
This choice can be rationalized by the physical picture that at lower densities, electrons prefer to stay further away from each other. 
Furthermore, this correlator, regardless of the choice of $k_c$, retains the cusp condition for two electrons with unlike-spins
at $r=0$~\cite{Luo2018} and should increase the convergence rate of the computed energies with respect to the 
employed basis set towards the CBS limit.

\begin{figure}[th!]
  \centering
  \includegraphics[width=\linewidth]{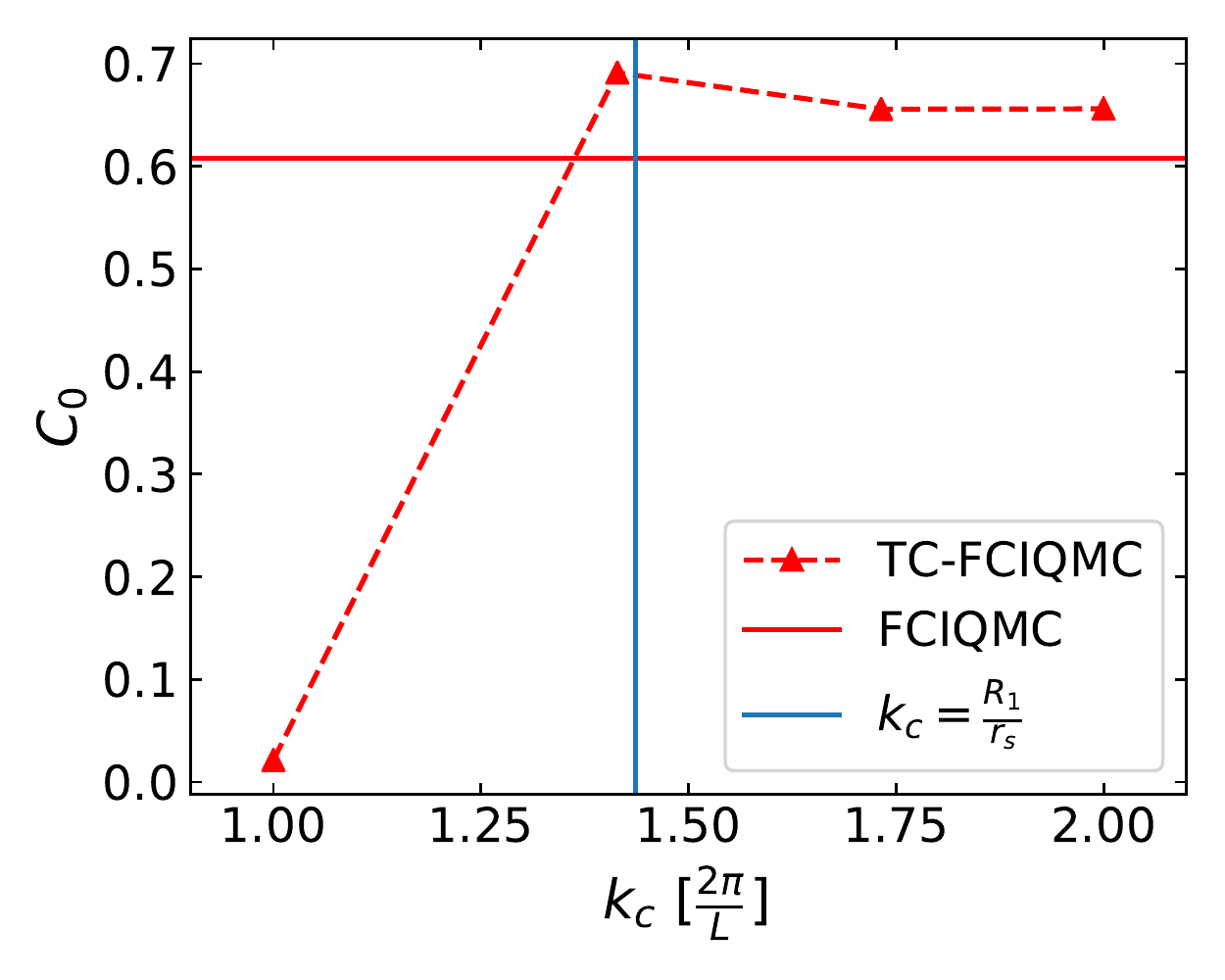}
  \caption{
      The weights of the HF determinant (Fermi sphere) as a function of $k_c$ extracted
    from their corresponding TC-FCIQMC simulations (red dashed line), and
    that of the normal FCIQMC simulation (horizontal red solid line).
    The system consists of 
    14 electrons with $r_s=5$ and a basis set including 57 plane waves.
    $5\times10^8$ walkers are used in all simulations and the initiator threshold is set
    to 3. 
    No approximations are made to the 3-body interactions.
  }
  \label{fig:HFCoefficient}
\end{figure}

To further justify the choice of this correlator, we show that for UEG with 14 electrons at
$r_s=5$, where traditional CCD exhibits a large error,
the most compact expansion
of the wavefunction in Slater determinant space is reached at this value of $k_c$.
In Fig.~\ref{fig:HFCoefficient}, we show
the weights of the HF determinant extracted from  TC-FCIQMC simulations using 
different $k_c$ values, without
making approximations to the 3-body operator.
We note that due to the discrete momentum mesh as a result of using
a finite simulation cell, the possible choices are 
$k_c=\frac{2\pi\sqrt{n}}{L}$, $n\in \mathbb{N}$~\footnote{$n=n_1^2+n_2^3+n_3^2$, where $n_1,n_2,n_3$ are the components of a ${\bf k}$ vector in units of $\frac{2\pi}{L}$. So the possible values of n are 0, 1, 2, 3, 4, 5, 6, 8, ...},
where $L$ is the length of the cubic cell. In this case, $k_c=\frac{R_1}{r_s}$ is
equivalent to $k_c=\frac{2\sqrt{2}\pi}{L}$, and for this choice of $k_c$ the exact ground state wavefunction of the transcorrelated Hamiltonian has the highest weight on the
HF determinant. 

\begin{figure}[th!]
  \centering
  \subfigure[\label{fig:14e-t2norm-scan}]{\includegraphics[width=0.5\linewidth]{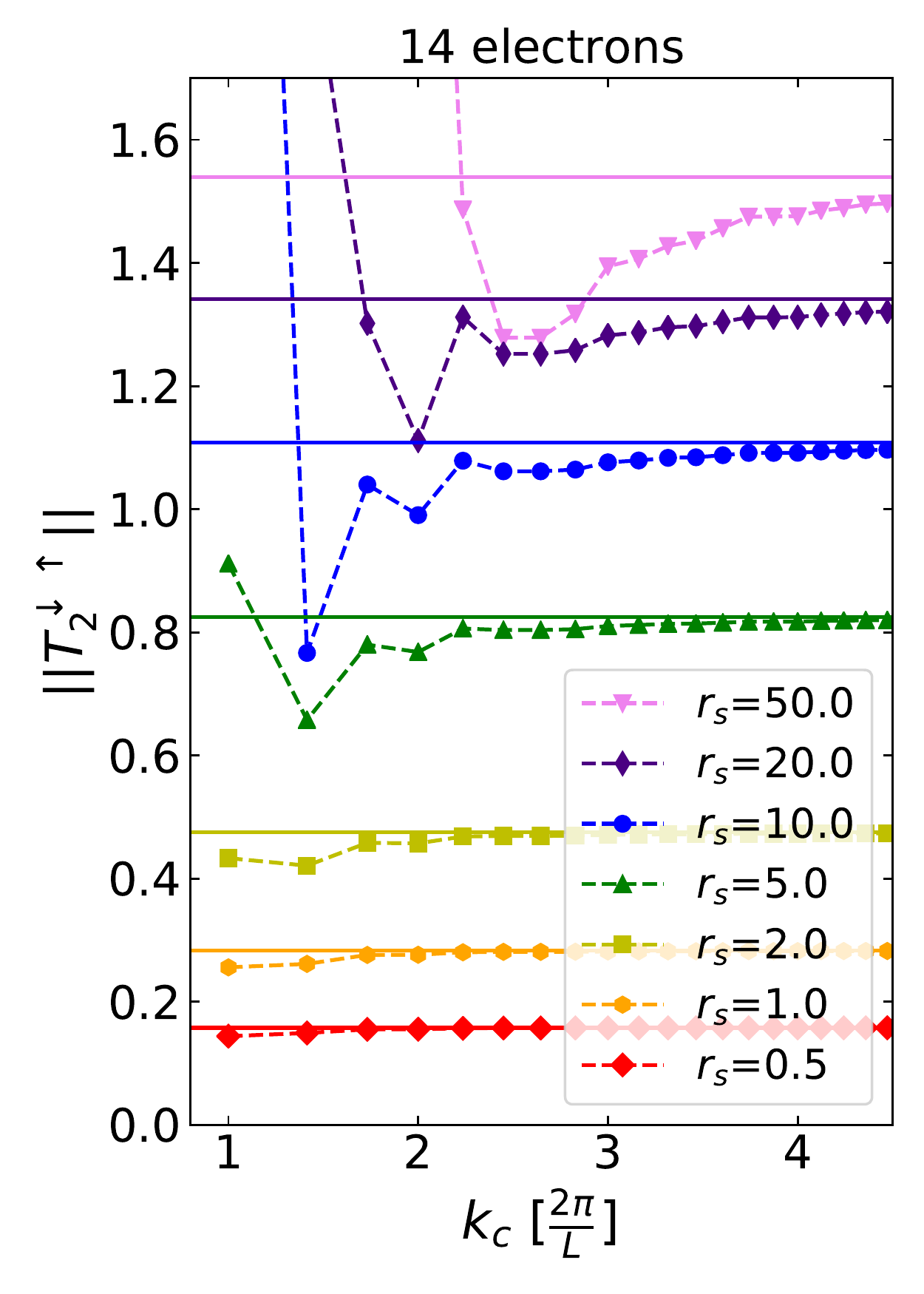}}%
  \subfigure[\label{fig:54e-t2norm-scan}]{\includegraphics[width=0.5\linewidth]{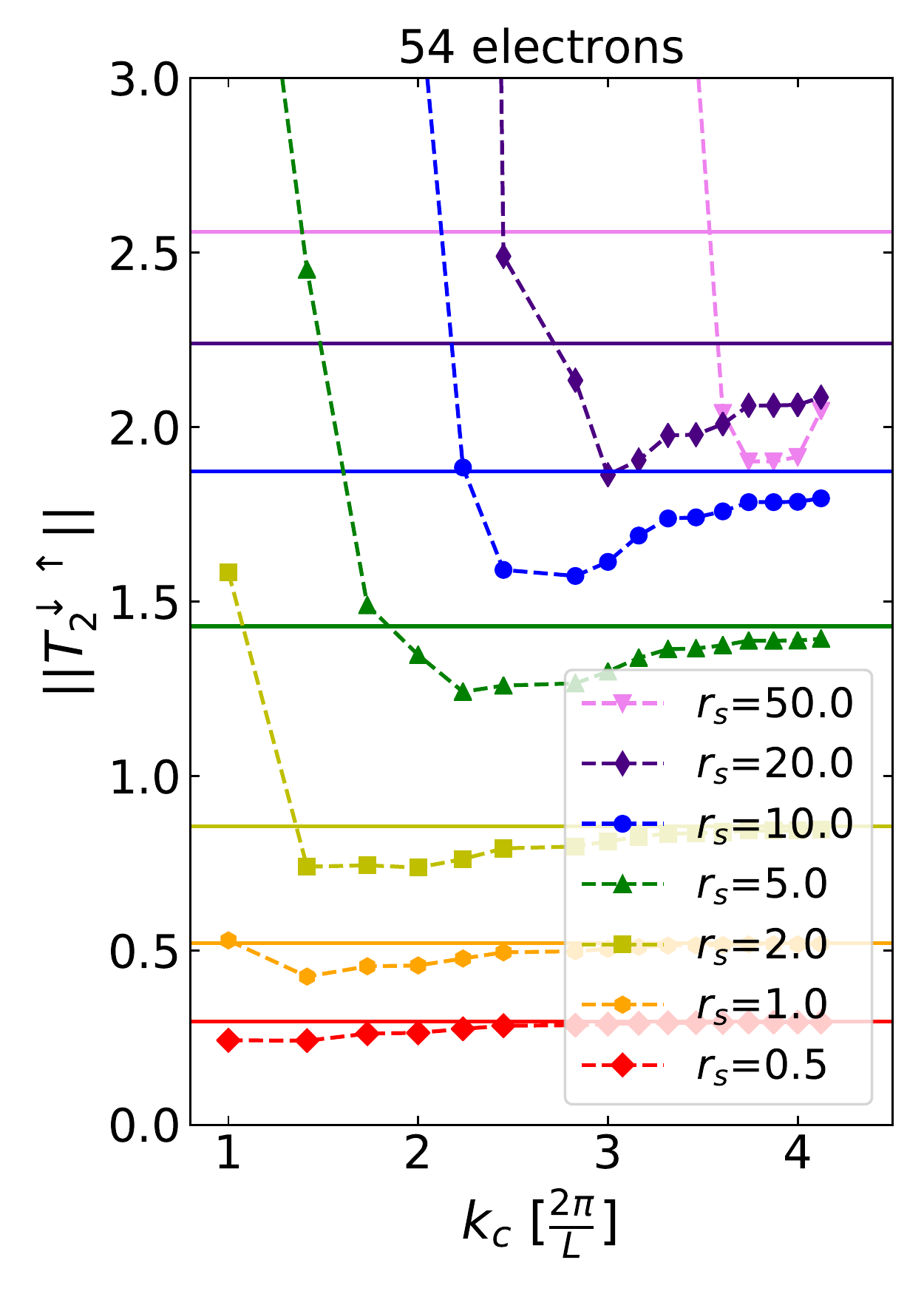}}%
  \caption{
    The norm of the amplitudes for excitations of unlike-spin electron pairs, $\|T_2^{\downarrow\uparrow}\|$, as a function of $k_c$, calculated
    by TC-DCD method. 
    The 14-electron and 54-electron 
    systems use a basis set including 57 and 257 plane waves, respectively.
    All possible contractions from the 3-body interactions are included, excluding the
    normal-ordered 3-body interactions.
    The solid horizontal color lines represent the $\|T_2^{\downarrow\uparrow}\|$ in the canonical 
    DCD calculations with the same settings as their TC counterparts.
  }
  \label{fig:t2norm_tc_dcd}
\end{figure}

However, we find that this intuitive choice of $k_c$ is not always the optimal, especially at extremely low density regimes. 
It is reasonable to expect that the optimal $k_c$ for those systems 
should deviate slightly from $\frac{R_1}{r_s}$. So 
we scan a range of $k_c$ values around it to locate the one that 
minimizes the norm of the closed-shell amplitudes for double excitations of electrons with opposite spins, $\|T_2^{\downarrow\uparrow}\|$, in the TC-CCD/DCD calculations with a small basis set, see Fig.~\ref{fig:t2norm_tc_dcd}~\footnote{Since the 
$\|T_2^{\downarrow\uparrow}\|$ in TC-CCD and TC-DCD show the same trend as a function of $k_c$, we show only the latter in the figure.}.

Ideally, two separate correlators should be used for electrons with parallel and anti-parallel spins, and their parameters should be optimized simultaneously using the norm of the full amplitudes in a similar manner. For the present study, we argue that the correlations between two parallel-spin electrons are dominated by the exchange effects, which are already captured by the anti-symmetry in the Slater
determinants.
Therefore, we focus on capturing the correct physics between electrons with opposite spins in the correlator, i.e.~the changing depth and width of the correlation hole as a function of $r_s$~\cite{Gori-Giorgi2000}, and minimizing the corresponding amplitudes in the CC ans\"atze. Indeed, we found in practice the minima in $\|T_2^{\downarrow\uparrow}\|$ as a function of $k_c$ are more pronounced, and thus easier to spot than those in the norm of the full amplitudes, $\|T_2\|$. 
We stress that this compact form of wavefunction at the optimal choice of $k_c$ 
should greatly benefit  approximate methods like
CCD and DCD, whose accuracy relies on the assumption that the true ground
state wavefunction is compact around the reference determinant, which is normally  chosen
to be the HF determinant.

%By avoiding choosing a very small $k_c$ value when $r_s$ is large and 
By including the most important contractions,
the error of neglecting the rest of the 3-body interactions are well under control in that it scales
approximately as $\tilde{u}^2(k)k^2\sim\frac{1}{k^6}$.
%In practice, if a too small $k_c$ is used,  The same happens 
We note in passing that if we choose correlators
that do not truncate at small $k$, such as the Yukawa-Coulomb correlator in Ref.~\cite{Gruneis2013b}
or the Gaskell correlator in Ref.~\cite{Gaskell1961}, the iterative solution of the amplitude equations  becomes too unstable to converge 
at low densities. 
We attribute this instability in these cases to the large missing normal-ordered 3-body interactions,
similar to the instability in a HF self-consistent solution when the missing normal-ordered
Coulomb interactions are large.

The FCIQMC calculations are carried out using the NECI program~\cite{Guther2020}.
The CCD and DCD along with the TC integrals are implemented in a Python program using
the automatic tensor contraction engine CTF~\cite{Solomonik2014a} and the NumPy package~\cite{Harris2020}.

\section{Results on the 3D UEG}
\label{sec:Results}
\begin{table*}[t]
\begin{tabular}{cccccccc}
\hline
\hline
  $r_s$ & $k_c$ ($\frac{2\pi}{L}$) & CCD & DCD & TC-CCD & TC-DCD & TC-FCIQMC & BF-DMC \\
\hline
0.5  & 1           & 3.41278  & 3.41252  & 3.41258  & 3.41244  & 3.41241(1)  & 3.41370(2)   \\
1.0  & 1           & 0.56975  & 0.56909  & 0.56891  & 0.56859  & 0.56861(1)  & 0.56958(1)     \\
2.0  & $\sqrt{2}$  & -0.00623 & -0.00748 & -0.00707 & -0.00800 & -0.00868(2) & -0.007949(7)       \\
5.0  & $\sqrt{2}$  & -0.07618 & -0.07788 & -0.07816 & -0.07929 & -0.08002(2) & -0.079706(3)  \\
10.0 & $\sqrt{2}$  & -0.05137 & -0.05289 & -0.05420 & -0.05509 & N/A         & -0.055160(2) \\
20.0 & 2           & -0.02924 & -0.03035 & -0.03136 & -0.03201 & N/A         & -0.0324370(8)\\
50.0 & $\sqrt{6}$  & -0.01261 & -0.01323 & -0.01350 & -0.01384 & N/A         & -0.0146251(3)\\
\hline
\hline
\end{tabular}
\caption{Total energy (a.u./electron), including the Madelung constant, of the 14-electron 3D UEG using different methods.}
\label{table:14e}
\end{table*}

\begin{table*}[t]
\begin{tabular}{cccccccc}
\hline
\hline
  $r_s$ & $k_c$ ($\frac{2\pi}{L}$) & CCD & DCD & TC-CCD & TC-DCD & TC-FCIQMC & BF-DMC \\
\hline
0.5  & $\sqrt{2}$   & 3.22079  & 3.22052  & 3.22077  & 3.22071  & 3.22042(2) & 3.22112(4)   \\
1.0  & $\sqrt{2}$   & 0.53069  & 0.53001  & 0.52982  & 0.52968  & 0.52973(3) &  0.52989(4)     \\
2.0  &  2           & -0.01162 & -0.01286 & -0.01324 & -0.01379 & N/A        &  -0.01311(2)     \\
5.0  & $\sqrt{5}$   & -0.07492 & -0.07655 & -0.07750 & -0.07837 & N/A        &  -0.079036(3)\\
10.0 &  $2\sqrt{2}$ & -0.05016 & -0.05157 & -0.05230 & -0.05322 & N/A        &  -0.054443(2) \\
20.0 &  3           & -0.02846  & -0.02925 & -0.03055 & -0.03113 & N/A        & -0.032047(2)\\
50.0 &  4           & -0.01223 & -0.01267 & -0.01263 & -0.01281 & N/A        & N/A \\
\hline
\hline
\end{tabular}
\caption{Total energy (a.u./electron), including the Madelung constant, of the 54-electron 3D UEG using different methods.}
\label{table:54e}
\end{table*}
\begin{figure}[th!]
  \centering
  \subfigure[\label{fig:14e-bsc-rs05}]{\includegraphics[width=0.5\linewidth]{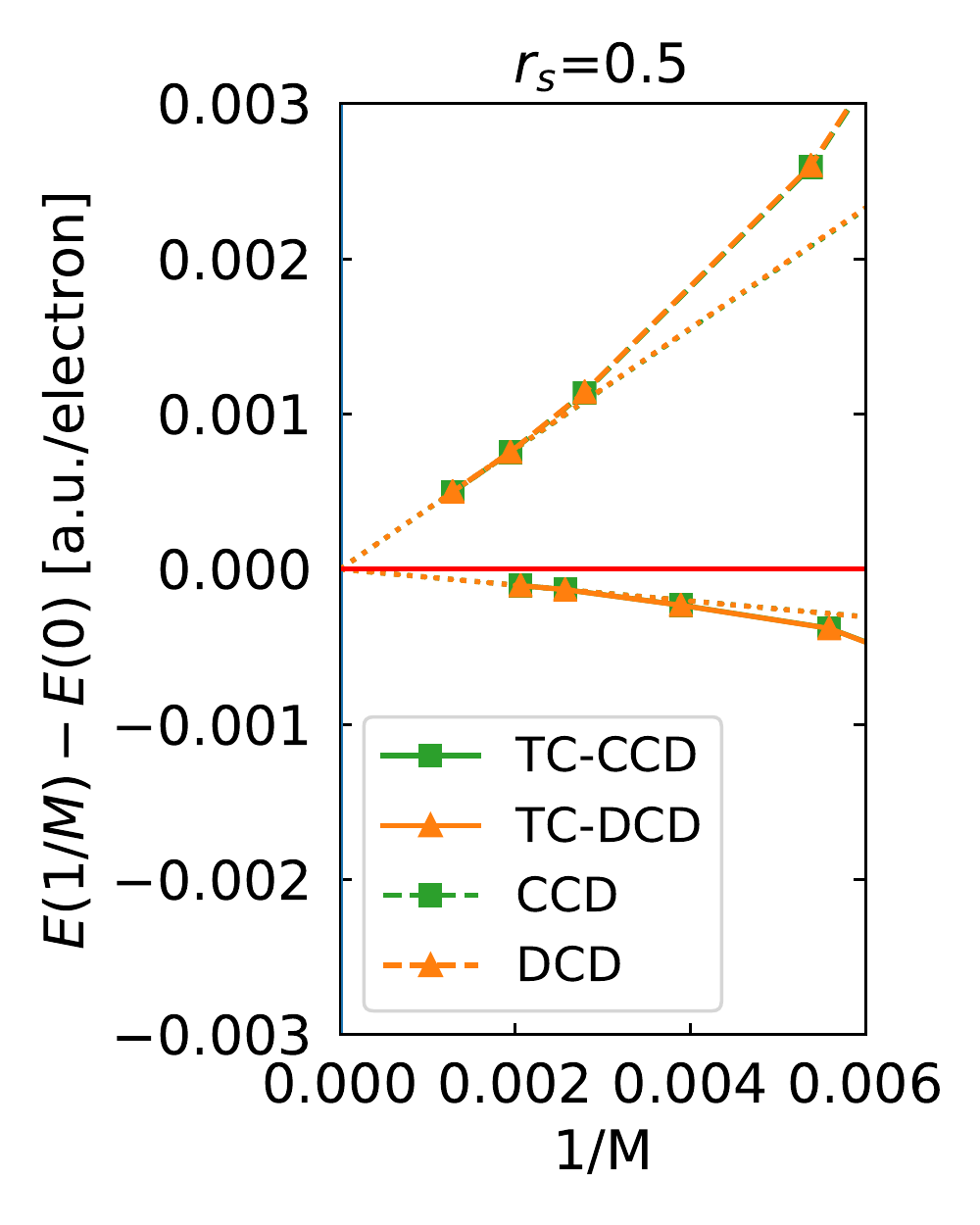}}%
  \subfigure[\label{fig:14e-bsc-rs50}]{\includegraphics[width=0.5\linewidth]{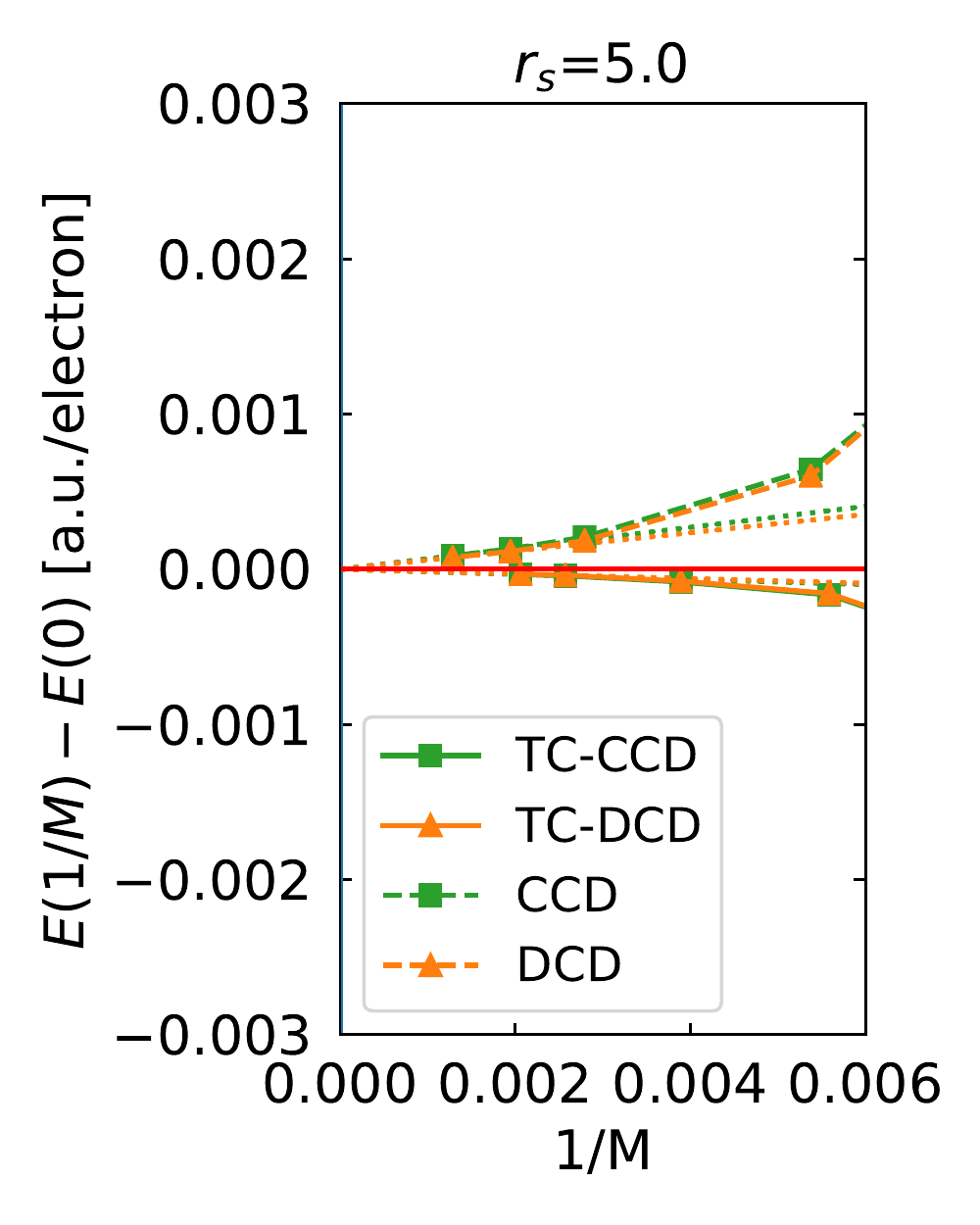}}%
  \caption{Total energy per electron relative to the extrapolated value retrieved as a function
  of $1/M$, where $M$ is the number of plane wave basis functions for 3D UEG with 14 electrons. 
  Linear extrapolations using the left most two points are used in all cases.
  }
  \label{fig:bsc}
\end{figure}

\begin{figure*}[th]
  \centering
  \subfigure[\label{fig:14e-error}]{\includegraphics[width=0.45\linewidth]{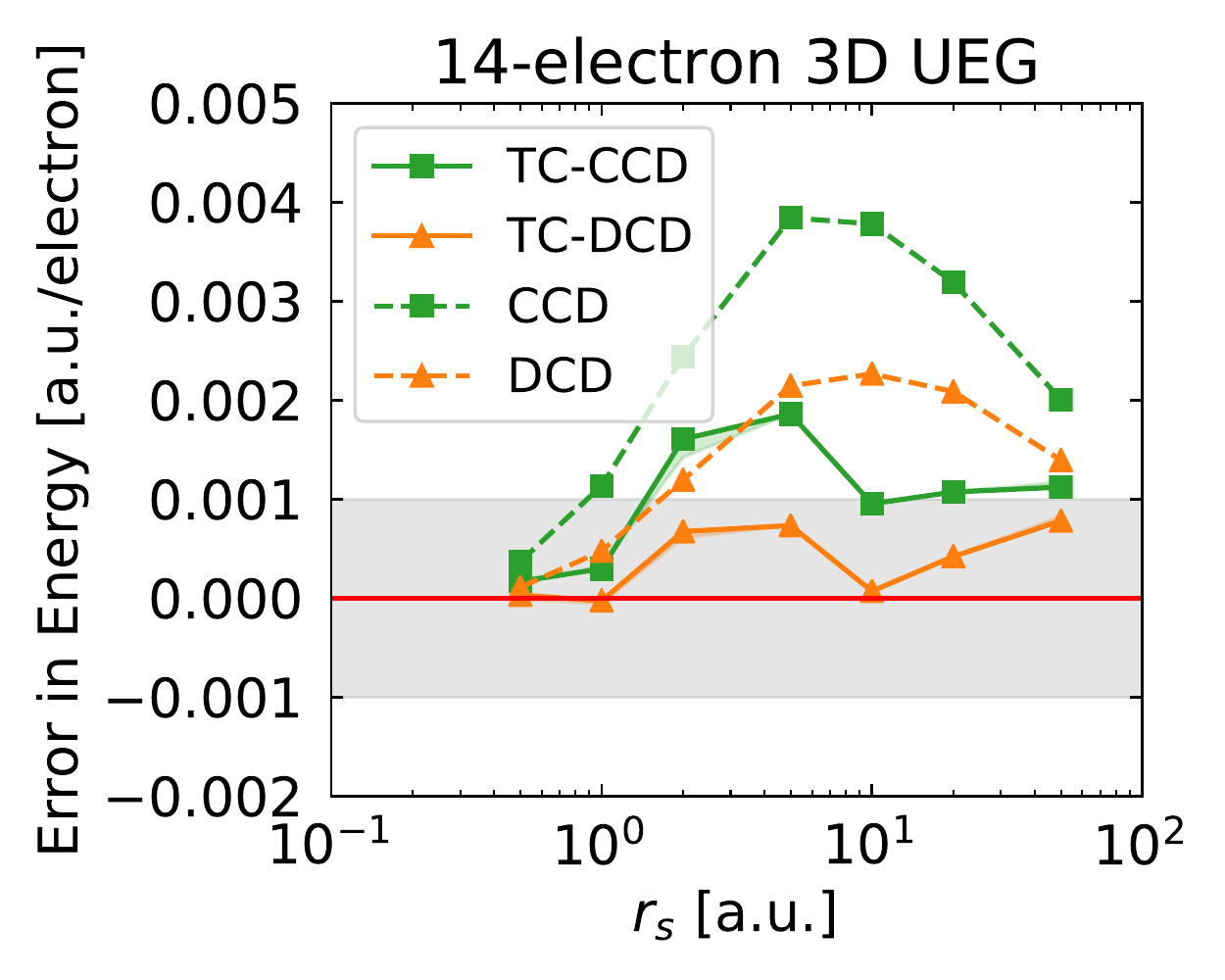}}%
  \subfigure[\label{fig:54e-error}]{\includegraphics[width=0.45\linewidth]{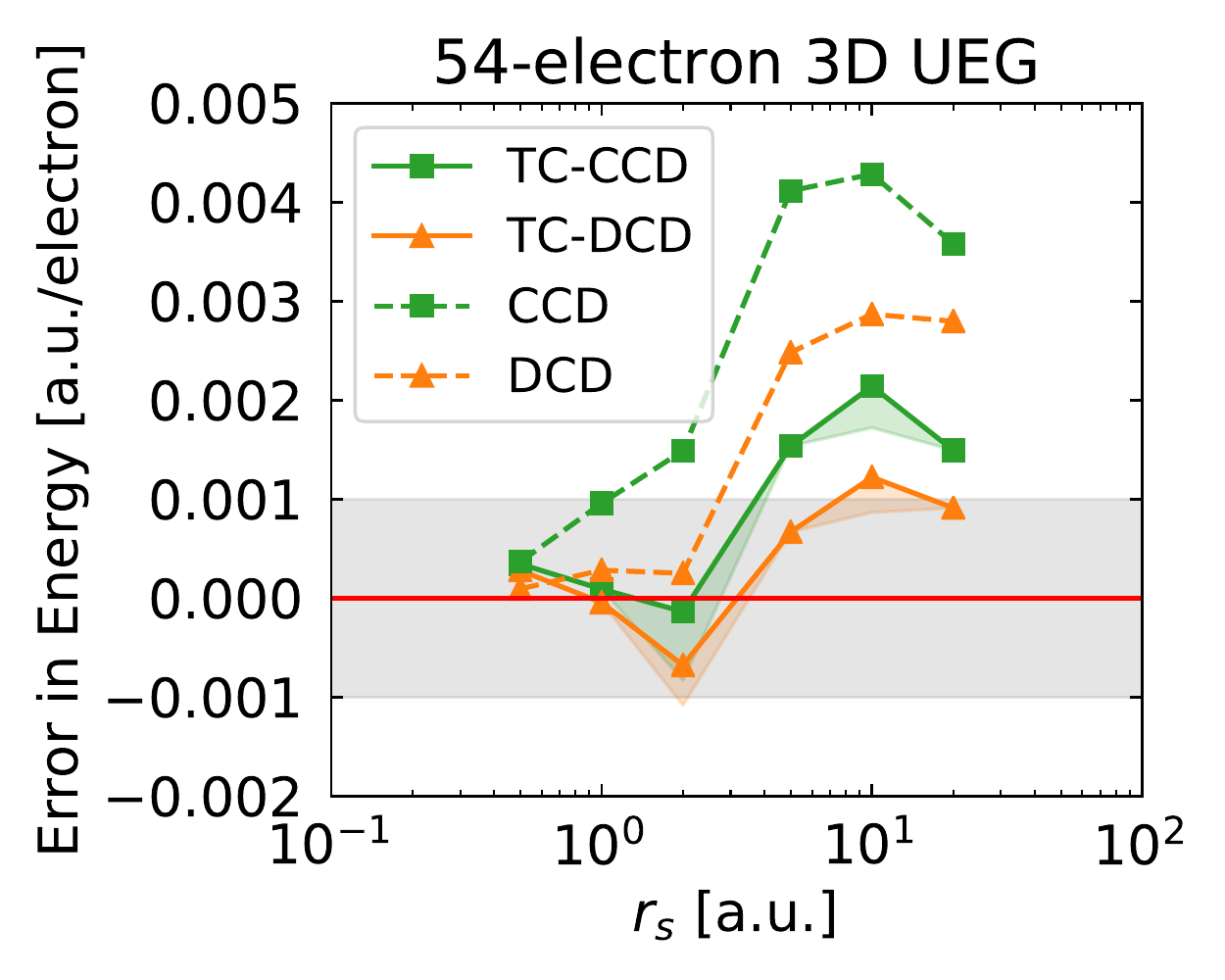}}
  \caption{Errors in energy per electron relative to benchmark data on the 3D UEG using
  TC-CCD, TC-DCD, CCD and DCD methods. 
  (a) Results for 14 electrons. For $r_s=0.5-5$, 
  TC-FCIQMC data~\cite{Luo2018} are used and 
  for the rest BF-DMC data~\cite{Shepherd2014a}
  are used for benchmark.
  (b) Results for 54 electrons. For $r_s=0.5-1$, TC-FCIQMC data~\cite{Luo2018} are used and for the rest BF-DMC 
  data~\cite{LopezRios2006} are used for benchmark.
  The grey shaded areas stand for the $\pm$0.001 a.u./electron accuracy
  relative to the reference data.
  The colorful shaded areas reflect the uncertainties in the
  TC-CCD and TC-DCD energies due to slightly different choices of the
  $k_c$ values.
  }
  \label{fig:tcresults}
\end{figure*}

We first examine the basis set convergence  behavior of TC-CCD/DCD compared
to the canonical ones. In Fig.~\ref{fig:bsc}
we present the total energy per electron relative to the extrapolated value
for each method, 
retrieved as a function of the inverse of the employed number of plane waves, $1/M$.
As mentioned before, our correlator satisfies the cusp condition at the coalescence point of two electrons with opposite spins. So the 
accelerated convergence behavior in the TC methods compared to the canonical ones is not 
surprising.
The acceleration is the most obvious at high density regimes, since at low densities
the required number of basis functions to reach convergence in both the TC and non-TC methods
is relatively small. 
These observations are consistent with those of the TC-FCIQMC reported in Ref.~\cite{Luo2018}.

The optimal $k_c$ values, (TC-)CCD/DCD energies at CBS and the benchmark data
are listed in Table~\ref{table:14e} and~\ref{table:54e} for the 14- and 54-electron
3D UEG, respectively.
In Fig.~\ref{fig:tcresults} we present the errors of total energies per electron
calculated by TC-CCD, TC-DCD, CCD and DCD relative
to the most accurate FCIQMC~\cite{Luo2018,Shepherd2012h,Neufeld2017} 
and backflow DMC (BF-DMC)~\cite{LopezRios2006,Shepherd2014a} results
on the 14- and 54-electron 3D UEG.
The finite basis set errors in our methods have been carefully eliminated by extrapolation
to the CBS limit.

In general the accuracies of the TC methods are greatly improved compared to their canonical
counterparts, especially in regions ($r_s=5-50$) where the latter exhibit the largest errors. More
importantly, the improved accuracies are retained when going from the 14- to the 54-electron
system. We highlight that the TC-DCD achieves an accuracy of $\leq 0.001$
a.u./electron
across a wide range of densities, i.e.~$r_s=0.5-50$ for the 14-electron and $r_s=0.5-20$ for the
54-electron 3D UEG, with
an exception at $r_s=10$ for the latter where it drops slightly out of
the 0.001 a.u./electron accuracy. We argue that with the next possible smaller value of $k_c=\sqrt{6}$, which
yields a marginally lower $\|T_2^{\downarrow\uparrow}\|$, instead of the current choice of $k_c=2\sqrt{2}$,
the 0.001 a.u./electron accuracy at $r_s=10$ can be regained. The discrete grid of the $k$-mesh makes it hard to pick the optimal $k_c$ in 
Fig.~\ref{fig:54e-t2norm-scan}. However, as the system gets larger and the
$k$-mesh gets finer, the $\|T_2^{\downarrow\uparrow}\|$ as a function of
$k_c$ will also be smoother, and the choice of the optimal $k_c$ will become more definite. We use colorful shaded areas in Fig.~\ref{fig:tcresults}
to reflect the uncertainties due to the possible choices of $k_c$ which
yield similar $\|T_2^{\downarrow\uparrow}\|$ values in Fig.~\ref{fig:t2norm_tc_dcd}.

At high densities, i.e.~$r_s=0.5-2$, the
canonical DCD is already very accurate, and the main benefit from TC there is in accelerating the basis set convergence.
Overall, 
DCD exhibits smaller errors than CCD,
which agrees with earlier comparative studies between DCSD and CCSD~\cite{Kats2013c,Kats2014,kats_accurate_2015}. 
%The DCD/DCSD neglects exchange 
%interactions between pairs of clusters, while keeps those within
%clusters. The idea of discarding small and unphysical exchange interactions 
%between possibly far apart pairs of electrons clearly works also in the 3D UEG, especially
%at low densities where electrons prefer to stay away from each other.

\section{Conclusions}
\label{sec:Con}
We demonstrated that the correlator Eq.~(\ref{equ:ReCorrelator}), used with transcorrelated coupled cluster theory,
drastically improves the accuracy of approximate
methods, i.e.~CCD and DCD, for 3D UEG across a wide range of densities.
The basis set convergence rate is also improved thanks to the fact that the correlator
satisfies the cusp condition at the coalescence point of two unlike-spin electrons. We have explored the mechanism 
behind the improved accuracy of the TC-CCD and TC-DCD methods, which is related to a
compactification of the many-electron wavefunction in Slater determinant space
when the dominant pair correlations between electrons 
with unlike spins are directly included in the correlator.
The optimization of the parameter in the correlator is seamlessly incorporated within
the TC-CCD/DCD framework, without requiring an external algorithm.
We notice
that a range-separation scheme of CCD can also achieve similar accuracy in 3D UEG, but without
improving the basis set convergence rate~\cite{Shepherd2014a}. Comparatively speaking, our methods
are systematically improvable, in that 
a more flexible form of the correlator can be
designed by a combination of a series of functions~\cite{Cohen2019} 
or by a general function approximator, e.g.~an artificial neural network, to include
further correlation effects such as nuclear-electron correlations and
correlations between parallel pairs of electrons. 
Other systematic ways of optimizing the correlator
in combination with VMC~\cite{Cohen2019} can also be explored.
When going to real periodic solids, TC-CCSD and TC-DCSD will be needed;
extra efforts are also required to compute the additional integrals besides the Coulomb integrals,
where the most computationally demanding part is the singly-contracted 
3-body integrals which scales like $\mathcal{O}(N_o^2N_v^4)$, where $N_o$ 
and $N_v$ are the number of occupied and unoccupied orbitals, 
respectively. Fortunately, the computation of the extra integrals scales no worse than the CCSD or DCSD algorithm
and it can
be compensated by the accelerated convergence rate towards CBS limit in the TC
framework.
These perspectives will be important in extending the
encouraging performance of the current TC-CCD and TC-DCD methods from the UEG to real
periodic solids with moderate to strong correlation.

{\bf Competing Interests} The authors declare that they have no
competing financial interests.

{\bf Acknowledgments} We thank A.~Gr\"uneis and N.~Masios for helpful
discussions and P.~L\'{o}pez R\'{i}os for providing the BF-DMC data for the 14 electron UEG.

\bibliography{arxiv_version}

\end{document}